\documentclass{nejsds}\usepackage[]{graphicx}\usepackage[dvipsnames,table]{xcolor}
\makeatletter
\def\maxwidth{ %
  \ifdim\Gin@nat@width>\linewidth
    \linewidth
  \else
    \Gin@nat@width
  \fi
}
\makeatother

\definecolor{fgcolor}{rgb}{0.345, 0.345, 0.345}
\newcommand{\hlnum}[1]{\textcolor[rgb]{0.686,0.059,0.569}{#1}}%
\newcommand{\hlsng}[1]{\textcolor[rgb]{0.192,0.494,0.8}{#1}}%
\newcommand{\hlcom}[1]{\textcolor[rgb]{0.678,0.584,0.686}{\textit{#1}}}%
\newcommand{\hlopt}[1]{\textcolor[rgb]{0,0,0}{#1}}%
\newcommand{\hldef}[1]{\textcolor[rgb]{0.345,0.345,0.345}{#1}}%
\newcommand{\hlkwb}[1]{\textcolor[rgb]{0.69,0.353,0.396}{#1}}%
\newcommand{\hlkwc}[1]{\textcolor[rgb]{0.333,0.667,0.333}{#1}}%
\newcommand{\hlkwd}[1]{\textcolor[rgb]{0.737,0.353,0.396}{\textbf{#1}}}%

\usepackage{framed}
\makeatletter
\newenvironment{kframe}{%
 \def\at@end@of@kframe{}%
 \ifinner\ifhmode%
  \def\at@end@of@kframe{\end{minipage}}%
  \begin{minipage}{\columnwidth}%
 \fi\fi%
 \def\FrameCommand##1{\hskip\@totalleftmargin \hskip-\fboxsep
 \colorbox{shadecolor}{##1}\hskip-\fboxsep
     \hskip-\linewidth \hskip-\@totalleftmargin \hskip\columnwidth}%
 \MakeFramed {\advance\hsize-\width
   \@totalleftmargin\z@ \linewidth\hsize
   \@setminipage}}%
 {\par\unskip\endMakeFramed%
 \at@end@of@kframe}
\makeatother

\definecolor{shadecolor}{rgb}{.97, .97, .97}
\definecolor{messagecolor}{rgb}{0, 0, 0}
\definecolor{warningcolor}{rgb}{1, 0, 1}
\definecolor{errorcolor}{rgb}{1, 0, 0}
\newenvironment{knitrout}{}{} 

\usepackage{alltt}

\volume{0}
\issue{0}
\pubyear{2022}
\articletype{research-article}
\doi{0000}

\usepackage{amsmath, amssymb} 
\usepackage{array} 
\usepackage{booktabs} 
\usepackage[dvipsnames,table]{xcolor}
\definecolor{lightgray}{gray}{0.95} 
\usepackage{url}
\usepackage{hyperref}

\IfFileExists{upquote.sty}{\usepackage{upquote}}{}
\begin{document}

\begin{frontmatter}

\pretitle{Research Article}

\title{Bayes Factor Group Sequential Designs}

\begin{aug}
\author[a]{\inits{S.}\fnms{Samuel} \snm{Pawel}\thanksref{c1}\ead[label=e1]{samuel.pawel@uzh.ch}}

\author[b]{\inits{L.}\fnms{Leonhard} \snm{Held}\ead[label=e2]{leonhard.held@uzh.ch}}
\thankstext[type=corresp,id=c1]{Corresponding author.}

\address[a]{Epidemiology, Biostatistics and Prevention Institute (EBPI), Center for Reproducible Science and Research Synthesis (CRS), \institution{University of Zurich}, \cny{Switzerland}.\\ \printead{e1}}

\address[b]{Epidemiology, Biostatistics and Prevention Institute (EBPI), Center for Reproducible Science and Research Synthesis (CRS), \institution{University of Zurich}, \cny{Switzerland}.\\ \printead{e2}}
\end{aug}

\begin{abstract}
The Bayes factor, the data-based updating factor from prior to posterior odds, is a principled measure of relative evidence for two competing hypotheses. It is naturally suited to sequential data analysis in settings such as clinical trials and animal experiments, where early stopping for efficacy or futility is desirable. However, designing such studies is challenging because computing design characteristics, such as the probability of obtaining conclusive evidence or the expected sample size, typically requires computationally intensive Monte Carlo simulations, as no closed-form or efficient numerical methods exist. To address this issue, we extend results from classical group sequential design theory to sequential Bayes factor designs. The key idea is to derive Bayes factor stopping regions in terms of the \textit{z}-statistic and use the known distribution of the cumulative \textit{z}-statistics to compute stopping probabilities through multivariate normal integration. The resulting method is fast, accurate, and simulation-free. We illustrate it with examples from clinical trials, animal experiments, and psychological studies. We also provide an open-source implementation in the \texttt{bfpwr} R package. Our method makes exploring sequential Bayes factor designs as straightforward as classical group sequential designs, enabling experiments to rapidly design informative and efficient experiments.
\end{abstract}


\end{frontmatter}

\section{Introduction}

A crucial decision in the design of an experiment is choosing the sample size,
for example, the number of participants, animals, or measurements. A too small
sample size risks yielding inconclusive results, whereas a too large sample size
may be too costly, logistically challenging, or ethically problematic.
Sequential designs have often been proposed as a means to reduce sample size
\citep{Kairalla2012}. In a sequential design it is possible to stop an
experiment early on when conclusive evidence is found in an interim analysis.
For example, a clinical trial may be stopped early if there is evidence that a
treatment is effective (stopping for efficacy) or if there is evidence that a
treatment is ineffective or even harmful (stopping for futility).

Despite their appeal, sequential designs pose statistical and practical
challenges. For example, they can bias parameter estimates \citep{Robertson2022}
or threaten the integrity of a blinded trial, as interim analyses may require
preliminary unblinding \citep{Ellenberg2019}. Furthermore, the analysis of
frequentist sequential designs can be unintuitive. For instance, a final
\textit{p}-value below the conventional 5\% threshold may still be insufficient
to declare efficacy if it does not cross a more stringent threshold due to
repeated analyses. Many practitioners find this difficult to accept, as the same
data would permit an efficacy claim if only one analysis had been conducted
\citep{Matthews2006}.

\subsection{Bayesian approaches to sequential analysis}

Bayesian methods are often considered as more natural for sequential analyses
\citep{JackLee2012}. Repeated ``updating'' is inherent in the
Bayesian framework and Bayesian probabilities are, in principle, not affected by
multiplicity issues \citep[however, see][]{Ryan2020, Zhou2023}. For this reason,
regulatory agencies are increasingly receptive to Bayesian approaches, and the
FDA has recently issued draft guidance explicitly permitting Bayesian methods
for primary inference in clinical trials, provided they are appropriately
justified \citep{FDA2026, Lee2026}.

Two popular Bayesian approaches to sequential analysis are based on posterior
tail probabilities \citep{berry2010bayesian, Gsponer2013, Shi2019, Rosner2020}
and Bayes factors \citep{Schoenbrodt2018, Johnson2009, Li2017, Pourmohamad2022}.
For directional hypotheses (e.g., testing whether a treatment effect is less or
greater than zero), the two are closely related, as a Bayes factor stopping rule
is equivalent to a posterior tail probability rule with a correspondingly
transformed threshold depending on the prior odds. Bayes factors, however, more
naturally accommodate point (or ``sharp'') null hypotheses, such as the absence
of a treatment effect, since the Bayes factor quantifies the evidence for such a
hypothesis without requiring a prior probability to be assigned to it. As point
null hypotheses are central to confirmatory and regulatory testing, Bayes
factors are particularly appealing in clinical trial settings.

Bayes factors also possess another property attractive to regulators. The
expected weight of evidence (the expected log Bayes factor) is largest when the
prior under the alternative matches the true data-generating mechanism, so
specifying an overly optimistic alternative prior can only diminish the expected
evidence for the treatment rather than bias the trial toward it
\citep{Johnson2009}. This contrasts with inference based on posterior tail
probabilities, where an optimistic prior can bias conclusions in favor of the
treatment, which is a commonly encountered argument against using Bayesian
methods \citep{JackLee2012}.

Bayes factors are also naturally suited to sequential analysis. Under the null
hypothesis, the Bayes factor in favor of the alternative is a test martingale
\citep{Shafer2011}, so that the probability of ever obtaining misleading
evidence is bounded irrespective of how often the data are analyzed. This is in
contrast to standard posterior tail probabilities where an increasing number of
interim analyses inflates the probability of erroneously rejecting the null
hypothesis to unity (``sampling to a foregone conclusion'' \citep{Cornfield1966,
  Cornfield1966b}, see also Chapter 13.3.2 in \citep{Jennison2025}). In this
sense, Bayes factors can be viewed as Bayesian generalizations of likelihood
ratios \citep{Wald1947, Royall1997} and are closely connected to recently
popularized \textit{e}-value and ``anytime-valid'' methods \citep{Ramdas2023,
  Grunwald2024}.

\subsection{Sequential Bayes factor designs}

We focus on sequential Bayes factor designs, which have been applied across a
wide range of domains, including biomedical research \citep{Cornfield1966b,
  Cornfield1976, Spiegelhalter2004b, Goodman2005, Johnson2009, Li2017, Zhu2019,
  Zhou2021, Rosner2021, Moerbeek2021, Pourmohamad2022, Linde2023}, the social
sciences \citep{Schoenbrodt2017, Schoenbrodt2018, Stefan2019, Mani2021,
  Stefan2022, Stefan2024}, and A/B testing in the tech industry \citep{Deng2016,
  Lindon2022}. Central to such designs is the Bayes factor
\begin{align*}
  \mathrm{BF}_{01}
  = \underbrace{\frac{\Pr(H_0 \mid \text{data})}{\Pr(H_1 \mid \text{data})} \, \bigg/ \frac{\Pr(H_0)}{\Pr(H_1)}}_{\substack{\text{data-based} \\ \text{updating factor}}}
  = \underbrace{\frac{p(\text{data} \mid H_0)}{p(\text{data} \mid H_1)}}_{\substack{\text{relative predictive} \\ \text{performance}}},
\end{align*}
the data-based updating factor of the prior odds of two hypotheses $H_0$ and
$H_1$ to the corresponding posterior odds. This update is dictated by the ratio
of the data's marginal likelihood under each hypothesis. The hypothesis which
better predicts the data receives more support in terms of the Bayes factor
\citep{Kass1995}. Bayes factors thus provide a direct and interpretable measure
of relative evidence for $H_0$ and $H_1$, which, unlike posterior probabilities,
does not depend on the prior probabilities of $H_0$ and $H_1$.

In a sequential Bayes factor design, one fixes two Bayes factor thresholds $k_0
> 1$ and $k_1 < 1$, the thresholds for $H_0$ and $H_1$, respectively. Common
choices are symmetric thresholds, such as $k_0 = 1/k_1 = 3$ or $k_0 = 1/k_1 =
10$, corresponding to ``substantial'' and ``strong'' relative evidence under
Jeffreys' conventions \citep{Jeffreys:1961}. The experiment is then stopped for
either $H_0$ or $H_1$ as soon as the Bayes factor exceeds the corresponding
threshold. If $H_0$ and $H_1$ represents the absence and presence of an effect,
these rules naturally accommodate stopping for futility and efficacy,
respectively. Sequential Bayes factor designs thus naturally link the stopping
decision to a relevant statistical evidence measure.

\begin{figure}[!htb]
\begin{knitrout}
\definecolor{shadecolor}{rgb}{0.969, 0.969, 0.969}\color{fgcolor}

{\centering \includegraphics[width=\maxwidth]{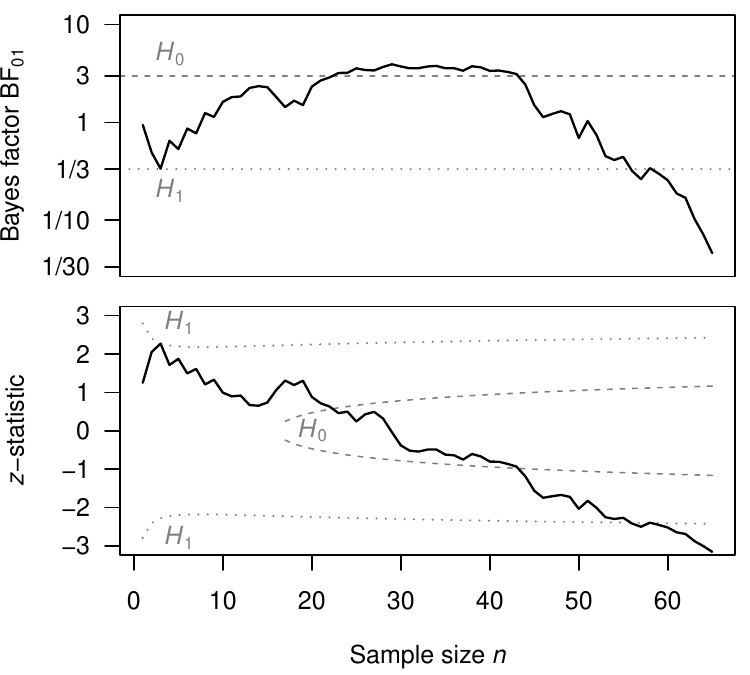} 

}

\end{knitrout}
\caption{Illustration of a sequential Bayes factor analysis. The Bayes factor
  quantifying the evidence for a point null hypothesis $H_0 \colon \theta = 0$
  against an alternative hypothesis $H_1 \colon \theta \neq 0$ is monitored as
  data accumulate (top). The bottom plot shows the corresponding $z$-statistic
  along with the Bayes factor stopping boundaries.}
\label{fig:bfcartoon}
\end{figure}

The top plot in Figure~\ref{fig:bfcartoon} shows a simulated trajectory of a
Bayes factor (oriented in favor of $H_0$ over $H_1$). Initially, the Bayes
factor fluctuates around 1, indicating absence of evidence for either
hypothesis. After around 20 observations, the Bayes factor surpasses the
threshold of 3, indicating substantial evidence for $H_0$ over $H_1$. However,
after around 50 observations, the Bayes factor quickly decreases below 1/3 and
1/10 -- as it should since the data were simulated under the alternative
hypothesis $H_1$. Depending on when the interim analyses were performed, the
experiment would thus have been erroneously stopped for $H_0$ or correctly
stopped for $H_1$.

The previous example illustrates the delicate choices that must be made when
setting up a sequential Bayes factor design. These include choosing Bayes factor
thresholds $k_0$ and $k_1$, deciding when to perform interim analyses, and
determining the maximum sample size. Exploring these options can be challenging
because computing design characteristics is typically only possible through
Monte Carlo simulation methods. For example, the method implemented in the
\texttt{BFDA} R package \citep{Schoenbrodt2019} involves analyzing many
simulated datasets to determine how often an analysis yields correct or
misleading results. While flexible, these methods are time-consuming and subject
to Monte Carlo error. This limits the experimenter's freedom in prototyping and
comparing different designs, which may ultimately result in the use of
suboptimal designs. Such barriers may also help explain the limited adoption of
sequential Bayes factor designs; indeed, a recent systematic review found that
full Bayesian methods for sample size determination remain rarely used in
randomized clinical trials \citep{Marks2026}. Addressing this gap is the aim of
the present paper.

\subsection{Contributions and outline}

In this paper, we present an alternative perspective on sequential Bayes factor
designs that enables fast, simulation-free computation of their design
characteristics. The key idea is to express the Bayes factor as a function of
the \textit{z}-statistic (the standardized difference between the observed
parameter estimate and the null value) so that the stopping rules correspond to
regions in the space of accumulating \textit{z}-statistics (see the bottom plot
in Figure~\ref{fig:bfcartoon}). Building on and extending results from classical
group sequential design theory, the probabilities of these regions, and hence
the design characteristics, can then be computed deterministically through
multivariate normal integration. The approach applies whenever the Bayes factor
can be expressed as a function of the \textit{z}-statistic and the accumulating
\textit{z}-statistics follow, at least approximately, a canonical multivariate
normal distribution. We additionally provide an open-source implementation in
the R package \texttt{bfpwr}, which makes planning sequential Bayes factor
designs as straightforward as planning classical group sequential designs.


In the following Section~\ref{sec:characteristics}, we outline the general
calculations of sequential Bayes factor design characteristics. We then explain
a method for calculating design characteristics efficiently and without
simulation when Bayes factors can be expressed as a function of the
\textit{z}-statistic (Section~\ref{sec:zstatistic}). Applications to clinical
trials and animal experiments illustrate how the method can be used in practice
(Section~\ref{sec:applications}). Section~\ref{sec:ttest} outlines an extension
to a sequential version of the Bayes factor \textit{t}-test \citep{Gronau2020}.
The paper closes with concluding discussions, limitations, and opportunities for
future research (Section~\ref{sec:discussion}). Appendix~\ref{app:package}
illustrates usage of our R package \texttt{bfpwr}.

\section{Sequential Bayes factor design characteristics}
\label{sec:characteristics}
Suppose we want to plan a sequential Bayes factor design with $m$ analyses. At
each analysis $i = 1, \dots, m$, based on the $n_i$ observations available up to
that point, a Bayes factor is computed and denoted by $\mathrm{BF}_{01}^i$.
Suppose that thresholds $k_0 > 1$ and $k_1 < 1$ are used and the experiment is
stopped as soon as evidence for either $H_0$ ($\mathrm{BF}_{01} \geq k_0$) or
$H_1$ ($\mathrm{BF}_{01} \leq k_1$) is found. Four quantities are of central
interest to assess the usefulness of a design:

\begin{enumerate}
\item \textbf{The probability of conclusive evidence for the alternative
  hypothesis} $(\mathrm{BF}_{01} \leq k_1)$. If data are generated under $H_1$
  this corresponds to the probability of correct evidence for $H_1$ -- a
  Bayesian analogue of the frequentist power (true positive rate). If data are
  generated under $H_0$ this corresponds to the probability of misleading
  evidence for $H_1$ -- a Bayesian analogue of the false positive (type-I) error
  rate.
\item \textbf{The probability of conclusive evidence for the null hypothesis}
  $(\mathrm{BF}_{01} \geq k_0)$. If data are generated under $H_1$ this
  corresponds to the probability of misleading evidence for $H_0$ -- a Bayesian
  analogue of the false negative (type-II error) rate. If data are generated
  under $H_0$ this corresponds to the probability of correct evidence for $H_0$
  -- a Bayesian analogue of the true negative rate.
\item \textbf{The expected sample size at termination}. A lower expected sample
  size is desired to reduce the actual sample size of the experiment.
\item \textbf{The standard deviation (or variance) of the sample size at
  termination}. A lower standard deviation is desired to better anticipate the
  actual sample size of the experiment.
\end{enumerate}

The probability of finding evidence for $H_1$ can be decomposed
\begin{align*}
  \small
  &\Pr(\text{Evidence for} ~ H_1) \\
  &= \Pr(\mathrm{BF}_{01}^1 \leq k_1) 
  \\
  &+\, \Pr(k_1 < \mathrm{BF}_{01}^1 < k_0, \mathrm{BF}_{01}^2 \leq k_1) 
  \\
  &~\vdots \\
  &+\, \Pr(k_1 < \mathrm{BF}_{01}^1 < k_0, \dots, k_1 < \mathrm{BF}_{01}^{m-1} < k_0,  \mathrm{BF}_{01}^m \leq k_1), 
\end{align*}
where the first summand represents the probability to find evidence in analysis
1, the second in analysis 2, and so on. The probability of finding evidence for
$H_0$ can be similarly expressed as
\begin{align*}
  &\Pr(\text{Evidence for} ~ H_0) \\
  &= \Pr(\mathrm{BF}_{01}^1 \geq k_0) 
  \\
  &+\, \Pr(k_1 < \mathrm{BF}_{01}^1 < k_0, \mathrm{BF}_{01}^2 \geq k_0) 
  \\
  &~\vdots \\
  &+\, \Pr(k_1 < \mathrm{BF}_{01}^1 < k_0, \dots, k_1 < \mathrm{BF}_{01}^{m-1} < k_0,  \mathrm{BF}_{01}^m \geq k_0). 
\end{align*}
Finally, the expected sample size is
\begin{align*}
  &\mathrm{E}(n) \\
  &= n_1 \times \{\Pr(\mathrm{BF}_{01}^1 \leq k_1) + \Pr(\mathrm{BF}_{01}^1 \geq k_0)\} 
  \\
  &+\, n_2 \times  \{ \Pr(k_1 < \mathrm{BF}_{01}^1 < k_0, \mathrm{BF}_{01}^2 \leq k_1)  \\
  &\phantom{+ n_2 \times \{} + \Pr(k_1 < \mathrm{BF}_{01}^1 < k_0, \mathrm{BF}_{01}^2 \geq k_0)\}
  \\
  &~\vdots \\
  &+\, n_m \times \Pr(k_1 < \mathrm{BF}_{01}^1 < k_0, \dots, k_1 < \mathrm{BF}_{01}^{m-1} < k_0), 
\end{align*}
where each possible sample size is multiplied by the probability to stop at the
corresponding stage. Likewise $\mathrm{E}(n^2)$ can be computed, based on which
also the variance of the sample size can be obtained. Instead of the variance,
the coefficient of variation $\mathrm{CV}(n) =
\sqrt{\mathrm{Var}(n)}/\mathrm{E}(n)$ may also be of interest to compare the
variability of the sample size across designs with differing expected sample
sizes. Each of these quantities thus consists of a (weighted) sum of
analysis-wise stopping probabilities. These stopping probabilities are generally
hard to compute due to the accumulating data having a complex distribution with
dependence across analyses. For this reason, Monte Carlo simulation is typically
used for approximating design characteristics. However, in the following we will
show that by assuming a particular form of Bayes factor and data, fast and
simulation-free computation is possible.

\section{Bayes factors based on \textit{z}-statistics}
\label{sec:zstatistic}
When testing hypotheses related to an unknown parameter $\theta$, many types of
Bayes factors $\mathrm{BF}_{01}$ can be expressed as a function of the
\textit{z}-statistic \mbox{$z = (\hat{\theta} - \theta_0)/\sigma$}, where
$\hat{\theta}$ is an estimate of $\theta$, $\sigma$ is the estimate's standard
error (assumed to be known), and $\theta_0$ is the null value related to the
null hypothesis $H_0$ (typically $\theta_0 = 0$). Table~\ref{tab:BFcritical}
shows several Bayes factor types for which this is possible. These are related
to a general class of Bayes factors based on test statistics \citep{Johnson2005,
  HeldOtt2018}. Note that throughout this paper, at the design stage $\sigma$ is
treated as a known function of $n$ (typically of form $\sigma = \lambda /
\sqrt{n}$ with $\lambda^2$ a unit variance and $n$ the effective sample size),
as in classical group sequential design theory, whereas at analysis it is
replaced by the usual estimated standard error.

Assume now that we are interested in the critical \textit{z}-values(s) for which
$\mathrm{BF}_{01} = k$ for some $k > 0.$ For many Bayes factors in
Table~\ref{tab:BFcritical}, critical values can be derived in closed-form (see
the right column). Although closed-form solutions are convenient, they are not
necessary for computing design characteristics as described in the following
sections. For example, the Bayes factor that contrasts a point null hypothesis
($H_0 \colon \theta = 0$) with a directional alternative hypothesis ($H_1 \colon
\theta > 0$) (bottom row in Table~\ref{tab:BFcritical}) does not have an
analytically available critical \textit{z}-value. However, numerical
root-finding can be used to determine it, as the Bayes factor can be expressed
as a function of the \textit{z}-statistic.

\begingroup {\small \renewcommand{\arraystretch}{1.3} 
\begin{table*}[!htb]
  \centering
\caption{Different types of Bayes factors for data in the form of an estimate
  $\hat{\theta}$ of the parameter $\theta$ with (assumed to be known) standard
  error $\sigma$, which is assumed to be normally distributed $\hat{\theta} \mid
  \theta \sim \mathrm{N}(\theta, \sigma^2)$, and the critical values for the
  corresponding $z$-statistic $z = \hat{\theta}/\sigma$ so that the Bayes factor
  equals a threshold $\mathrm{BF}_{01} = k$.}
\label{tab:BFcritical}
\rowcolors{1}{}{gray!15}
\resizebox{1\linewidth}{!}{%
 \begin{tabular}{p{0.5\linewidth} p{0.5\linewidth}}
    \toprule
    \multicolumn{1}{c}{\textbf{Bayes factor}} & \multicolumn{1}{c}{\textbf{Critical \textit{z}-value(s)}} \\
    \midrule
    Directional null ($H_0 \colon \theta \leq 0$) vs. directional alternative ($H_1 \colon \theta >
    0$) with marginal normal prior $\theta \sim \mathrm{N}(\mu, \tau^2)$ \newline
    $$ \mathrm{BF}_{01} = \frac{1 - \Phi(\mu_*/\tau_*)}{\Phi(\mu_*/\tau_*)} \, \bigg / \, \frac{1 - \Phi(\mu/\tau)}{\Phi(\mu/\tau)}$$
    with $\tau_{*}^2 = 1/(1/\sigma^2 + 1/\tau^2)$ and $\mu_{*} = (z_i/\sigma +
\mu/\tau^2) \tau^2_{*}$
    & $$z_{\text{crit}}(k) = \left(\Phi^{-1}\left[\left\{k \, \frac{1 - \Phi(\mu/\tau)}{\Phi(\mu/\tau)} + 1\right\}^{-1}\right] \sqrt{\frac{1}{\sigma^2} + \frac{1}{\tau^2}} - \frac{\mu}{\tau^2}\right) \sigma$$ \\
    Point null ($H_0 \colon \theta = 0$) vs. point alternative ($H_1 \colon \theta = \mu$)
    \newline
    $$\mathrm{BF}_{01}
    = \exp\left(\frac{\mu^2}{2\sigma^2} - \frac{z \mu}{\sigma}\right)$$ & $$z_{\text{crit}}(k) =
    \frac{\mu^2/\sigma^2 - \log k^2}{2 \mu/\sigma}$$ \\
     Point null ($H_0 \colon \theta = 0$) vs. two-sided alternative ($H_1 \colon \theta \neq
    0$) with normal prior under alternative $\theta \mid H_1 \sim
    \mathrm{N}(\mu, \tau^2)$
    \newline
    $$\mathrm{BF}_{01} = \sqrt{1 + \frac{\tau^2}{\sigma^2}} \, \exp\left[-\frac{1}{2}\left\{z^2 -
    \frac{(z - \mu/\sigma)^2}{1 + \tau^2/\sigma^2} \right\}\right]$$
    & $$z_{\text{crit}-}(k) = M - \sqrt{X} ~~ \text{and} ~~
  z_{\text{crit}+}(k) = M + \sqrt{X}$$  \newline
  with $M = -\mu\sigma/\tau^2$ \newline and
  $X = \{\mu^2/\tau^2 + \log(1 + \tau^2/\sigma^2) - \log k^2\}(1 + \sigma^2/\tau^2)$
\\
    Point null ($H_0 \colon \theta = 0$) vs. two-sided alternative ($H_1 \colon \theta \neq
    0$) with normal moment prior under the alternative $\theta \mid H_1 \sim
    \mathrm{NM}(0, \tau^2)$ \newline
    $$\mathrm{BF}_{01} =
  \left(1 + \frac{\tau^{2}}{\sigma^{2}}\right)^{3/2} \, \exp\left\{\frac{-z^{2}}{2(1 + \sigma^{2}/\tau^{2})}\right\} \bigg /  \left(1 + \frac{z^{2}}{1 + \sigma^{2}/\tau^{2}}\right)$$ \newline
  A normal moment prior has density  $\mathrm{NM}(\theta \mid 0, \tau^2) = \mathrm{N}(\theta \mid 0, \tau^2) \times (\theta - 0)^2/\tau^2$ where $\mathrm{N}(\cdot \mid m, v)$ is the normal density function with mean $m$ and variance $v$ \citep{Johnson2010, Pramanik2024}
    & $$z_{\text{crit}-} = -\sqrt{Y} ~~ \text{and} ~~ z_{\text{crit}+} = +\sqrt{Y}$$  \newline
  with $Y = (2\mathrm{W}_0[\{ \sqrt{e}(1 + \tau^2/\sigma^2)^{3/2}\}/(2k)] - 1)(1 + \sigma^2/\tau^2)$ \newline
  and where $\mathrm{W}_0$ is the principal branch of the Lambert W function \citep{Corless1996} \\
Point null ($H_0 \colon \theta = 0$) vs. directional alternative ($H_1 \colon \theta >    0$) with truncated normal prior under alternative $\theta \mid H_1 \sim
    \mathrm{N}(\mu, \tau^2)_{(0, +\infty)}$ where the subscript denotes truncation of the
distribution to the interval $(0, +\infty)$
    \newline
    $$\mathrm{BF}_{01} = \sqrt{1 + \frac{\tau^2}{\sigma^2}} \, \exp\left[-\frac{1}{2}\left\{z^2 -
    \frac{(z - \mu/\sigma)^2}{1 + \tau^2/\sigma^2} \right\}\right] \, \frac{\Phi(\mu/\tau)}{\Phi(\mu_*/\tau_*)}$$ 
    & ~ \newline ~ \newline $z_{\text{crit}}$ not analytically available but can be determined \newline with numerical root-finding
\\
Point null ($H_0 \colon \theta = 0$) vs. directional alternative ($H_1 \colon \theta >
0$) with truncated normal moment prior under the alternative $\theta \mid H_1 \sim
\mathrm{NM}(0, \tau^2)_{(0, +\infty)}$ \newline
$$\mathrm{BF}_{01} =
\left(1 + \frac{\tau^{2}}{\sigma^{2}}\right)^{3/2} \bigg/ \left\{ 2 \exp\left(\frac{s^{2}}{2}\right) \left(1 + s^2\right) \Phi\left(s\right) +
  s \sqrt{\frac{2}{\pi}}\right\}$$ \newline
with $s = z/\sqrt{1 + \sigma^2/\tau^2}$
&  ~ \newline ~ \newline $z_{\text{crit}}$ not analytically available but can be determined \newline with numerical root-finding \\
    \bottomrule
  \end{tabular}
  }
\end{table*}
}
\endgroup

\begin{figure}[!htb]
\begin{knitrout}
\definecolor{shadecolor}{rgb}{0.969, 0.969, 0.969}\color{fgcolor}

{\centering \includegraphics[width=\maxwidth]{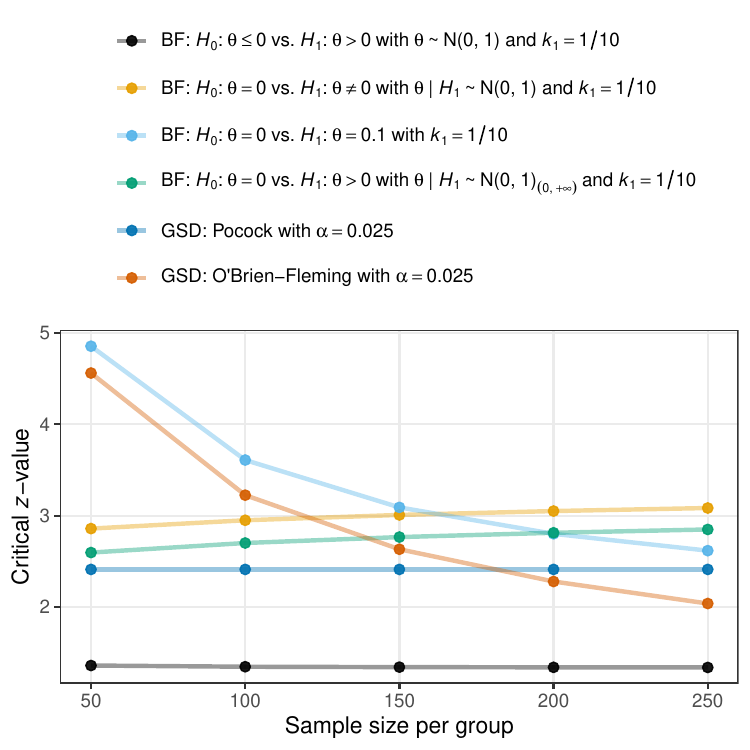} 

}

\end{knitrout}
\caption{Critical \textit{z}-values where $\mathrm{BF}_{01} = 1/10$
  for different types of Bayes factors (BF) from Table~\ref{tab:BFcritical} and
  group sequential designs (GSD). The tested parameter is the standardized mean
  difference between two normally distributed populations with know variance and
  standard error of the form $\sigma = \sqrt{2/n}$ where $n$ is the sample size
  per group.}
\label{fig:BFcritical}
\end{figure}

Figure~\ref{fig:BFcritical} shows criticial \textit{z}-values associated with
different Bayes factors indicating strong evidence for the alternative over the
null hypothesis ($\mathrm{BF}_{01} = 1/10$). The critical
\textit{z}-values from two commonly used group sequential design (GSD) methods,
Pocock and O'Brien-Fleming, are shown as comparison \citep[see e.g., Chapter~2
  in][]{Jennison1999}. Both ensure control of the type-I error rate at the
conventional $\alpha = 0.025$ (one-sided) in a sequential design with
5 equally-spaced analyses. Section~\ref{sec:canonical} will
describe how the type-I error rate of sequential Bayes factor designs can be
evaluated. We can see that the value and shape of the critical value boundaries
differs across the different Bayes factors. The directional null vs. directional
alternative Bayes factor ($H_0 \colon \theta \leq 0$ vs. $H_1 \colon \theta >
0$; black) shows near constant critical values across all analyses. In contrast,
the point null vs. point alternative Bayes factor ($H_0 \colon \theta = 0$ vs.
$H_1 \colon \theta = 0.1$; light-blue) shows decreasing critical values,
while the point null vs. directional alternative Bayes factor ($H_0 \colon
\theta = 0$ vs. $H_1 \colon \theta > 0$; green) shows increasing critical
values. The constant Pocock GSD critical values (dark-blue) thus align with the
shape of the directional null vs. directional alternative Bayes factor (black)
whereas the O'Brien-Fleming GSD critical values (orange) align with the shape
of the point null vs. point alternative Bayes factor (light-blue).

These resemblances should, however, not be mistaken for an equivalence. The
Bayes factor boundaries are determined by the evidence thresholds $k_0$ and
$k_1$ together with the hypotheses and priors, rather than calibrated to attain
a fixed type-I error rate, and they have a direct evidential interpretation.
Moreover, the point null vs. directional alternative (green) and the point null
vs. two-sided alternative (yellow) Bayes factors yield critical values that
increase with sample size which neither the Pocock nor the O'Brien-Fleming
boundary exhibits.

\subsection{Stopping regions based on \textit{z}-statistics}
\begin{figure*}[!htb]
\begin{knitrout}
\definecolor{shadecolor}{rgb}{0.969, 0.969, 0.969}\color{fgcolor}

{\centering \includegraphics[width=\maxwidth]{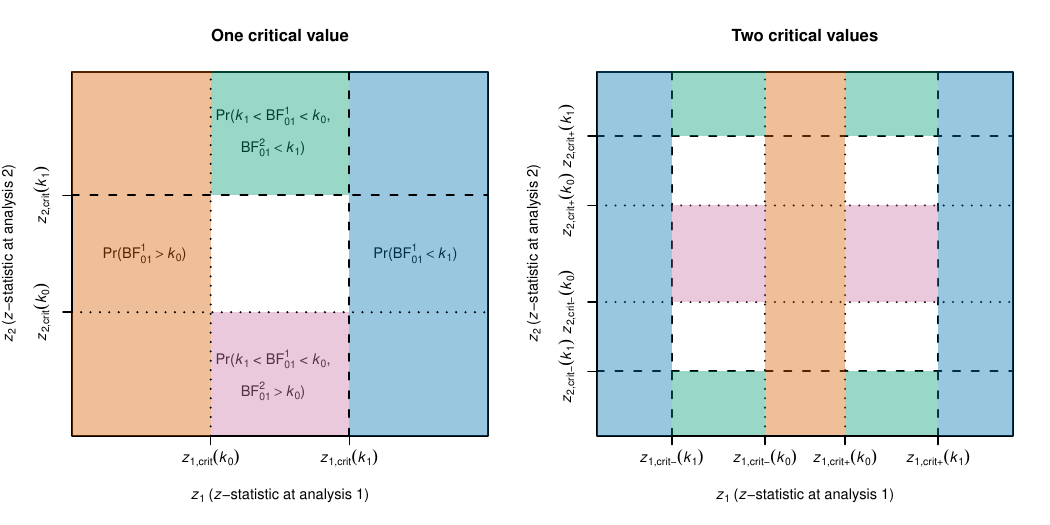} 

}

\end{knitrout}
\caption{Illustration of critical \textit{z}-values such that $\mathrm{BF}_{01}
  \geq k_0$ (dotted lines) and $\mathrm{BF}_{01} \leq k_1$ (dashed lines) for
  Bayes factors with one critical value (left) and two critical values (right)
  such that $\mathrm{BF}_{01} = k$. Integrating the colored regions produces the
  indicated analysis-wise stopping probabilitites. For example, integrating the
  green region gives the probability to not stop in the first analysis but stop
  for $H_1$ in the second analysis.}
\label{fig:2dz}
\end{figure*}

Denote by $z_i$ the \textit{z}-value at analysis $i$ and by
$z_{i,\text{crit}}(k)$ the critical value(s) for which the Bayes factor equals
the threshold $\mathrm{BF}_{01} = k$. For Bayes factors with only one critical
\textit{z}-value (e.g., directional null vs. directional alternative) the $H_0$
stopping condition $\mathrm{BF}_{01}^i \geq k_0$ is equivalent to $z_i \leq
z_{i,\text{crit}}(k_0)$ whereas the $H_1$ stopping condition $\mathrm{BF}_{01}^i
\leq k_1$ corresponds to $z_i \geq z_{i,\text{crit}}(k_1)$. The continuation
condition $k_1 < \mathrm{BF}_{01}^i < k_0$ is equivalent to
$z_{i,\text{crit}}(k_0) < z_i < z_{i,\text{crit}}(k_1)$. This means that the
\textit{z}-statistic stopping region at analysis $i$ for the vector of
cumulative \textit{z}-statistics $(Z_1, \dots, Z_i)^\top$ consists of an
$i$-dimensional rectangle (an ``$i$ hyper-rectangle''). Integrating this region
over the joint distribution of the $z$-statistics gives the probability to stop
at analysis $i$. For example, for $i = 1$ this is simply the interval
$[z_{1,\text{crit}}(k_1), +\infty)$ while for $i = 2$ we have the rectangle
  $(z_{1,\text{crit}}(k_0), z_{1,\text{crit}}(k_1)) \times
  [z_{2,\text{crit}}(k_1), +\infty)$, see the blue and green regions in the left
    plot in Figure~\ref{fig:2dz}.

For Bayes factors with two critical values (e.g., point null vs. two-sided
alternative), the conditions are more complicated. The $H_0$ stopping condition
$\mathrm{BF}_{01}^i \geq k_0$ is equivalent to the $z$-statistic being in the
interval around zero $z_i \in [z_{i,\text{crit}-}(k_0),
z_{i,\text{crit}+}(k_0)]$ whereas the $H_1$ stopping condition
$\mathrm{BF}_{01}^i \leq k_1$ corresponds to the $z$-statistic being outside the
interval $z_i \not\in (z_{i,\text{crit}-}(k_1), z_{i,\text{crit}+}(k_1))$. The
continuation condition $k_1 < \mathrm{BF}_{01}^i < k_0$ is equivalent to the
$z$-statistic being in between these two regions $z_i \in
(z_{i,\text{crit}-}(k_1), z_{i,\text{crit}-}(k_0)) \cup
(z_{i,\text{crit}+}(k_0), z_{i,\text{crit}+}(k_1))$.
This means that at analysis $i$, the stopping region consist of the union of
multiple $i$ hyper-rectangles, their number doubling with each analysis. For
example, in analysis $1$ we have two intervals of $z$-statistics for which the
Bayes factor is $\mathrm{BF}_{01} \leq k_1$, see the blue regions in the right
plot of Figure~\ref{fig:2dz}. In analysis 2, the number of rectangles doubles to
four (green rectangles in the right plot of Figure~\ref{fig:2dz}). This means
that with an increasing number of interim analyses, the number of
hyper-rectangles grows exponentially as every $i$ hyper-rectangle at analysis
$i$ splits into two $i + 1$ hyper-rectangles at analysis $i + 1$.

While the exponentially growing number of hyper-rectangles poses certain
computational challenges, we did not observe any numerical issues in our
applications when exhaustively computing the probability of all hyper-rectangles
in designs with even up to ten analyses. In many fields, designs involving such
a large number of analyses are rarely encountered. For example, the most common
number of interim analyses in clinical trials is only one or two, if any
\citep{Stevely2015}.
Moreover, for Bayes factors with only one critical \textit{z}-value (e.g.,
one-sided tests), the number of hyper-rectangles increases only linearly with
the number of analyses. Exhaustively computing their probabilities therefore
poses no computational problems, even for a very large number of analyses.

\subsection{The predictive distribution of the \textit{z}-statistics under a design prior}
\label{sec:canonical}
In order to compute the probability for the \textit{z}-statistics to fall within
a given region, we need to know their distribution. Suppose that $m$ analyses
produce the sequence of \textit{z}-statistics $\{Z_1, \dots, Z_m\}$. Under mild
conditions, these have an asymptotic \emph{canonical distribution}
\citep[Chapter~3]{Jennison1999}, that is, a distribution of the form
\begin{enumerate}
  \item $\boldsymbol{Z} = (Z_1, \dots, Z_m)^\top$ has an $m$-variate normal distribution
  \item $\mathrm{E}(\boldsymbol{Z}) = \theta \boldsymbol{I}$ where $\theta$ is
    the true parameter and $\boldsymbol{I} = (\sqrt{I_1}, \dots,
    \sqrt{I_m})^\top$ is the vector of square-rooted information levels. The
    information level is typically the inverse of the squared standard error
    $\sigma^2$. For example, for a normal mean, the information level is $I_i =
    1/\sigma^2 = n_i/\lambda^2$, where $n_i$ is the sample size at analysis $i$
    and $\lambda^2$ the known variance of one observation.
  \item $\mathrm{Cov}(Z_i, Z_{i+j}) = \boldsymbol{\Sigma}_{i,i+j} =
    \sqrt{I_i/I_{i+j}}$ for $j \geq 0$. For example, for a normal mean the
    covariance is $\mathrm{Cov}(Z_i, Z_{i+j}) = \sqrt{n_i/n_{i+j}}$.
\end{enumerate}
In short, the vector of \textit{z}-statistics has distribution
\begin{align*}
  \boldsymbol{Z} \mid \theta \sim \mathrm{N}_m\left(\theta \boldsymbol{I},
  \boldsymbol{\Sigma}\right).
\end{align*}

Assuming a fixed value for $\theta$, one can compute the probability of a
\textit{z}-statistic vector falling into a given stopping region using standard
numerical approaches for calculating multivariate normal integrals. Due to the
canonical distribution, this calculation can even be further simplified via a
recursive one-dimensional integration algorithm, which is also used extensively
in the calculation of stopping probabilities in classical group sequential
designs \citep{Armitage1969, Jennison1999}. However, from a Bayesian
perspective, it seems more natural to account for uncertainty of $\theta$ by
specifying a \emph{design prior} distribution \citep{OHagan2001b}, sometimes
also known as sampling prior \citep{wang2002, Psioda2018}, instead of a fixed
value. Taking the design prior to be a normal prior $\theta \sim
\mathrm{N}(\mu_d, \tau_d^2)$, the marginal (or prior-predictive) distribution of
$\boldsymbol{Z}$ is then
\begin{align}
  \label{eq:zmarginal}
  \boldsymbol{Z} \mid \mu_d, \tau_d^2 \sim \mathrm{N}_m\left(\mu_d \boldsymbol{I} ,
  \boldsymbol{\Sigma} + \tau_d^2 \boldsymbol{I} \boldsymbol{I}^\top\right),
\end{align}
see Appendix~\ref{app:zmarginal} for a derivation. The
distribution~\eqref{eq:zmarginal} reduces again to the canonical distribution
if a point prior at $\mu_d$ is assigned (i.e., if $\tau_d \downarrow 0$).
However, for non-degenerate priors ($\tau_d > 0$), the vector of accumulating
test statistics $\boldsymbol{Z}$ exhibits higher correlations than under the
canonical distribution, which increases with increasing $\tau_d$.

In contrast to the analysis prior, which is typically ``weakly-informative'' or
``objective'' in some sense, the design prior should represent genuine knowledge
and uncertainty at the design analysis in order to accurately estimate stopping
probabilities and sample sizes. Averaging operating characteristics over the
design prior in this way yields their Bayesian predictive analogues. For
instance, the probability of conclusive evidence for $H_1$ is a Bayes factor
analogue of the assurance or predictive power, generalizing classical power,
which conditions on a single parameter value \citep{OHagan2001b,
  Spiegelhalter1986}. However, setting the design prior to a point mass ($\tau_d
\downarrow 0$) allows us to study the frequentist operating characteristics of
the sequential design. For example, specifying a point prior at $\mu_d = 0$ and
computing the probability to find evidence for $H_1$ gives the frequentist
type-I error rate. Although this probability may be unrealistic from a Bayesian
perspective, showing that a sequential Bayes factor design is appropriately
calibrated (i.e., has a type-I error rate below a conventional level, e.g., 5\%)
may be required from certain stakeholders, such as regulatory authorities in
drug development \citep{FDA2010, Campbell2020}. Both frequentist and Bayesian
operating characteristics can thus be computed within the same framework,
differing only in the choice of design prior.

Assuming there are $m$ analyses, the probability of evidence for $H_i \in \{H_0,
H_1\}$ is
\begin{align*}
  &\Pr(\text{Evidence for}~ H_i) \\
  &~=
  \sum_{j=1}^m \Pr\{\boldsymbol{Z}_{1:j} \in \boldsymbol{S}_{j}^i \mid \mu_d \boldsymbol{I}_{1:j},
  (\boldsymbol{\Sigma} + \tau_d^2 \boldsymbol{I} \boldsymbol{I}^\top)_{1:j,1:j}\}
\end{align*}
where $1:j$ indicates indexing of the first $j$ elements and $\boldsymbol{S}_{j}^i$
is the set of $z$-statistic stopping regions for $H_i$ at analysis $j$. For example, assuming that $m = 3$ and that there
is only one critical value so that $\mathrm{BF}_{01} = k$ (e.g., a directional
null vs. directional alternative Bayes factor in Table~\ref{tab:BFcritical}), we
have the stopping regions for $H_1$
\begin{align*}
  \boldsymbol{S}_{1}^1 &= [z_{1,\text{crit}}(k_1), +\infty) \\
  \boldsymbol{S}_{2}^1 &= \big(z_{1,\text{crit}}(k_0), z_{1,\text{crit}}(k_1)\big)
     \times [z_{2,\text{crit}}(k_1), +\infty) \\
  \boldsymbol{S}_{3}^1 &= \big(z_{1,\text{crit}}(k_0), z_{1,\text{crit}}(k_1)\big)
     \times \big(z_{2,\text{crit}}(k_0), z_{2,\text{crit}}(k_1)\big) \\
  &\phantom{= (} \times [z_{3,\text{crit}}(k_1), +\infty)
\end{align*}
corresponding to hyper-rectangles in one-, two-, and three-dimensional space.
While under the canonical distribution, these probabilities could be expressed
as recursive one-dimensional integrals, this is no longer possible with design
priors where \mbox{$\tau_d > 0$} due to the distribution not being canonical
anymore. However, we have found that computation of even very large-dimensional
integrals (e.g., designs with 20 analyses, which is unrealistic in practice) is
still very efficient with modern implementations of the multivariate normal
distribution, such as the \texttt{mvtnorm} R package \citep{Genz2009}. Such
extreme designs will be demonstrated in Sections~\ref{sec:rat}
and~\ref{sec:ttest}.

\section{Applications}
\label{sec:applications}
We will now illustrate planning of sequential Bayes factor designs using case
studies from clinical trials and animal experiments.

\subsection{The Low-PV trial}

The Low-PV trial \citep{Barbui2021} assessed if the drug ropeginterferon alfa-2b
could help low-risk polycythaemia vera (Low-PV) patients keeping haematocrit
levels (the volume percentage of red blood cells in blood) within a safe range.
The study was designed to have three analyses after 50, 100, and 150 patients
had been followed-up, respectively. The design assumed response rates of $\pi_0
= 50\%$ and $\pi_1 = 75\%$ under
$H_1$ in the control and treatment groups, corresponding to an odds ratio of
$\mathrm{OR} = 3$. The study was stopped after the second
interim analysis because the group sequential stopping bounds for efficacy were
crossed. Table~\ref{tab:lowpv} summarizes the by-analysis results.

\begin{table*}[!htb]
  \centering
  \caption{Results from the Low-PV trial \citep{Barbui2021}. Shown are
    by-analysis sample sizes ($n$) and estimated response probabilities
    ($\hat{\pi}$) in control (subscript 0) and treatment groups (subscript 1),
    respectively, along with estimated odds ratio ($\widehat{\text{OR}}$) with
    corresponding 95\% confidence interval (CI), \textit{z}-value, and decision
    based on group sequential stopping bounds. The right-most column gives the
    Bayes factor contrasting $H_0 \colon \mathrm{OR} = 1$ to $H_1 \colon
    \mathrm{OR} = 3$ based on an approximately normal likelihood
    of the estimated log OR.}
  \label{tab:lowpv}
\begin{tabular}{lllllllll}
  \toprule
Analysis & $n_0$ & $n_1$ & $\hat{\pi}_0$ & $\hat{\pi}_1$ & $\widehat{\mathrm{OR}}$ (95\% CI) & $z$ & Decision & $\mathrm{BF}_{01}$ \\ 
  \midrule
1 & 26 & 24 & 57.7\% & 87.5\% & 5.1 (1.2 to 21.6) & 2.23 & Continue & 1/9.2 \\ 
  2 & 50 & 50 & 60.0\% & 84.0\% & 3.5 (1.4 to  9.0) & 2.60 & Stop (efficacy) & 1/27.9 \\ 
   \bottomrule
\end{tabular}

\end{table*}

Assuming a normal likelihood for the estimated log odds ratio $\hat{\theta} =
\log \widehat{\mathrm{OR}}$ and using the hypotheses specified by the trial
investigators, we can compute the Bayes factor from Table~\ref{tab:BFcritical}
contrasting the point hypotheses $H_0 \colon \theta = 0$ to $H_1 \colon \theta =
\log(3) \approx 1.1$. This leads to a Bayes
factor $\mathrm{BF}_{01}^1 = 1/9.2$ in the first analysis
and $\mathrm{BF}_{01}^2 = 1/27.9$ in the second. Had we used
a threshold of $k_1 = 1/10$, the study would have thus been stopped for $H_1$
after the second analysis, resulting in the same decision as was made by the
trial investigators based on a classical group sequential design.

Suppose the trial has not been carried out and we want to plan it using a
sequential Bayes factor design. Assuming allocation of $n$ patients to each
group, the approximate standard deviation of the estimated log odds ratio
$\hat{\theta}$ can be derived via the delta method to be $\sigma = \sqrt{1/\{n
  \pi_0 (1 - \pi_0)\} + 1/\{n \pi_1 (1 - \pi_1)\}}$. Once data are observed, it
is estimated through the usual standard error obtained from plugging in the
estimated rates $\hat{\pi}_0$ and $\hat{\pi}_1$ \citep{Bland2000}. Assuming a
normal distribution for $\hat{\theta}$ around the true log odds ratio $\theta$
with standard deviation $\sigma$, implies a canonical distribution for $z =
\hat{\theta}/\sigma$ with information level $I = 1/\sigma^2$. We can therefore
apply the results from Section~\ref{sec:zstatistic} to compute sequential Bayes
factor design characteristics.

\begin{figure*}[!b]
\begin{knitrout}
\definecolor{shadecolor}{rgb}{0.969, 0.969, 0.969}\color{fgcolor}

{\centering \includegraphics[width=\maxwidth]{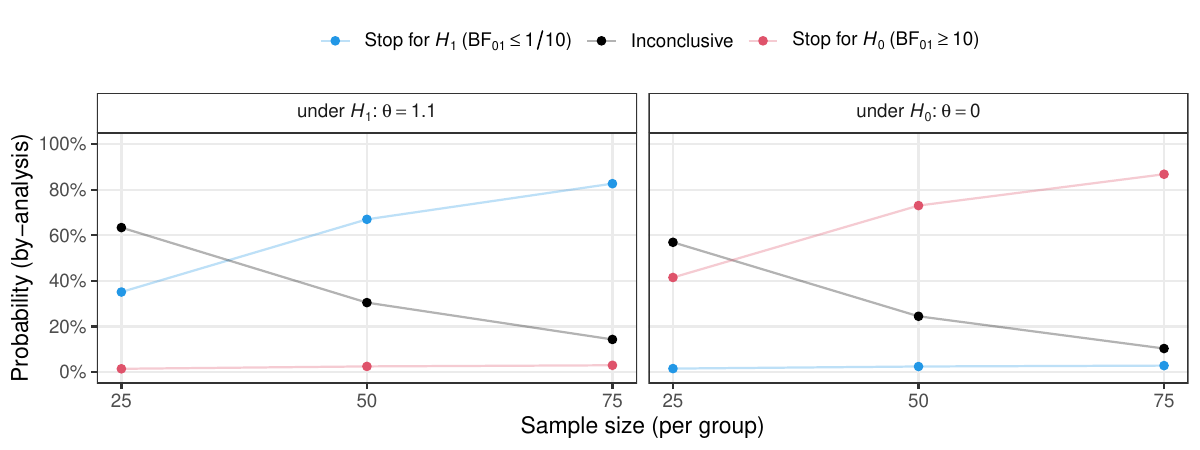} 

}

\end{knitrout}
\caption{Sequential Bayes factor probabilities for the Low-PV trial
  \citep{Barbui2021}.}
\label{fig:lowPVdesign}
\end{figure*}

Figure~\ref{fig:lowPVdesign} shows the probability of obtaining conclusive
evidence for $H_1$ (blue), $H_0$ (red), or to remain inconclusive (black) by
each of the three analyses, assuming that data are generated under $H_1$
($\theta = 1.1$; left plot) or under $H_0$ ($\theta =
0$; right plot). Three equally-spaced analyses at 25, 50, and 75 patients per
group are considered, as in the original trial. We can see that the
probabilities of conclusive evidence for the corresponding true hypotheses are
very similar under both hypotheses. Both surpass 80\% after the third analysis,
but remain below 90\%. To achieve a probability of 90\%, the maximum sample size
and/or the number of interim analyses must be further increased.

Keeping the number of analyses fixed at three, we can use numerical root-finding
(e.g., \texttt{uniroot} in R) to determine the maximum sample size size to
achieve a 90\% probability of conclusive evidence. This
leads to $n_3 = 87$ under $H_0$ and $n_3 =
102$ under $H_1$, translating into $29$
and $34$ additional patients per group at each analysis,
respectively. Taking the maximum of the two would ensure that the trial produces
conclusive evidence with high probability.

\begin{figure*}[!htb]
\begin{knitrout}
\definecolor{shadecolor}{rgb}{0.969, 0.969, 0.969}\color{fgcolor}

{\centering \includegraphics[width=\maxwidth]{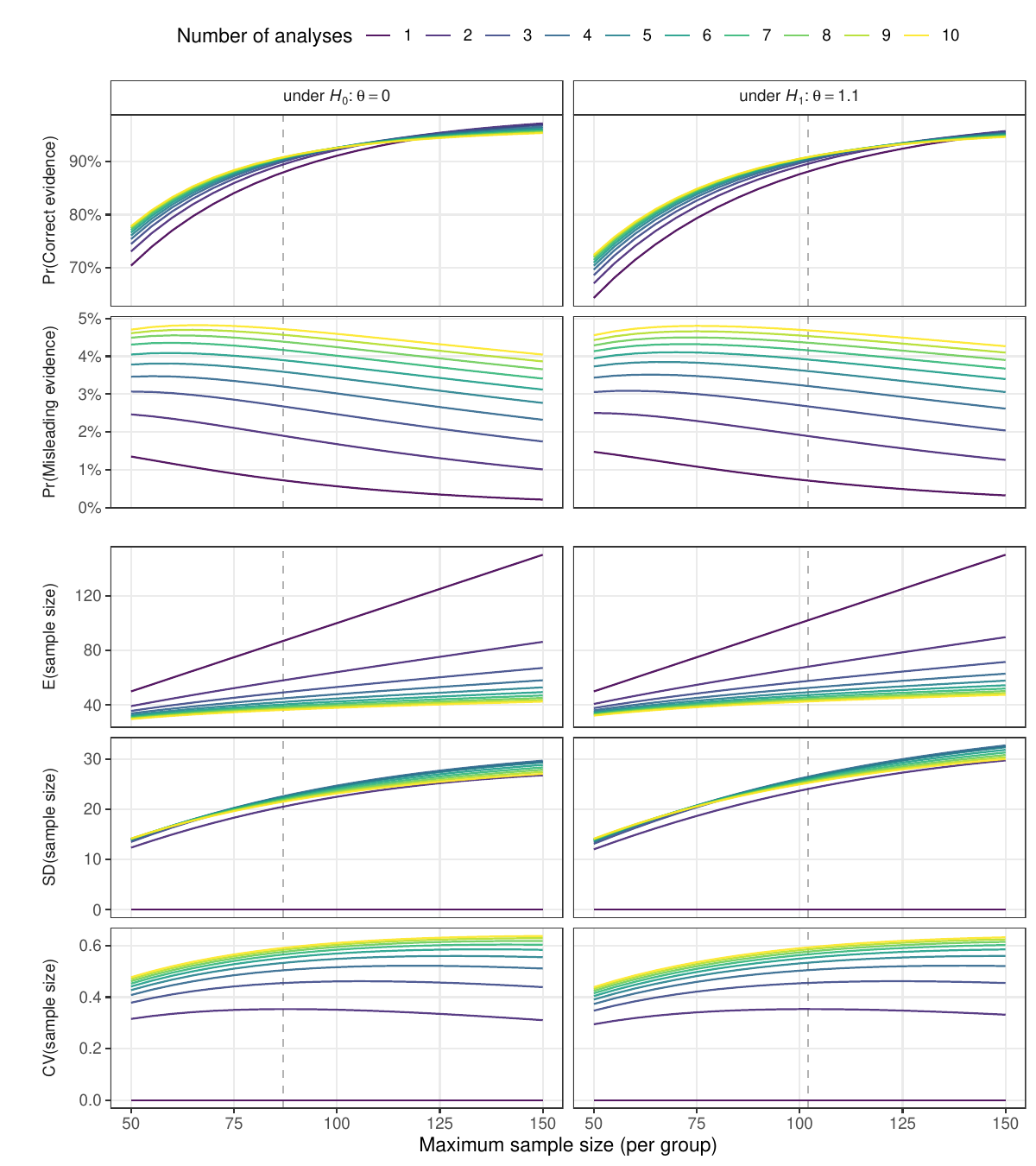} 

}

\end{knitrout}
\caption{Sequential Bayes factor design characteristics for the Low-PV trial
  \citep{Barbui2021} for differing maximum sample sizes and number of analyses.
  The dashed lines show the maximum sample sizes corresponding to
  90\% probability of correct evidence for 3
  analyses.}
\label{fig:lowPVdesign2}
\end{figure*}

Figure~\ref{fig:lowPVdesign2} shows design characteristics varying the maximum
samples size and also the numbers of analyses. This includes also only one
analysis which corresponds to a fixed design. Note that our approach can compute
these design characteristics in a few seconds, whereas simulating all these
designs would introduce Monte Carlo error and take substantially longer as for
every combination a sequential trial has to be simulated many times. From the
top-row plots, we can see that the probability of obtaining ``correct'' evidence
(i.e., $\mathrm{BF}_{01} \leq 1/10$, if $H_1$ is true and
$\mathrm{BF}_{01} \geq 10$, if $H_0$ is true) increases with increasing
maximum sample size (x axis). Increasing the number of analyses (color) also
increases the probability of obtaining correct evidence for low maximum sample
sizes. However, for maximum sample sizes higher than around 110, increasing the
number of analyses actually decreases the probability of obtaining correct
evidence. This is because a design with more analyses has also higher chances of
producing misleading evidence (i.e., $\mathrm{BF}_{01} \geq 10$ if $H_1$
is true and $\mathrm{BF}_{01} \leq 1/10$ if $H_0$ is true) and
stopping for the wrong hypothesis (second-row plots).

The probability of misleading evidence first increases and then decreases again
with increasing maximum sample size. Since $H_0$ and $H_1$ are point hypotheses,
the probability of misleading evidence is bounded by $k = 1/10$, which is known
as the ``universal bound'' \citep{Royall1997}. However, across all choices of
the number of analyses, the probability is much lower than the bound, remaining
below 5\% across all maximum sample sizes and number of analysises, though
bringing it below 2.5\% (the conventional type-I error rate for one-sided tests)
requires substantial increases in the maximum sample size for designs with many
analyses (not shown).

Increasing the number of analyses reduces the expected sample size (third-row
plots) due to the potential early stopping, while having a non-monotone effect
on the standard deviation of the sample size (fourth-row plots) -- increasing
for smaller maximum sample sizes and decreasing for larger maximum sample sizes.
Dividing the standard deviation by the expected sample size gives the
coefficient of variation (bottom-row plots), which enables sample size
variability to be compared while accounting for differing expected sample sizes.
For a given maximum sample size, the coefficient of variation increases with
increasing number of analyses, as more sample sizes at which stopping is
possible become available. At the same time, for a given number of analyses,
increasing the maximum sample size initially increases the coefficient of
variation until it reaches a maximum, after which it decreases again. However,
for a low number of analyses (e.g., two or three) this change is relatively
small so that the coefficient of variation remains almost constant.

In sum, compared to a fixed design, a sequential design with even only two or
three analyses can drastically improve efficiency. It reduces the expected
sample size and increase the probability of correct evidence, while only
slightly increasing the probability of misleading evidence and the variability
of the sample size.

\subsection{Rat experiment on weight loss}
\label{sec:rat}
\begin{figure}[!htb]
\begin{knitrout}
\definecolor{shadecolor}{rgb}{0.969, 0.969, 0.969}\color{fgcolor}

{\centering \includegraphics[width=\maxwidth]{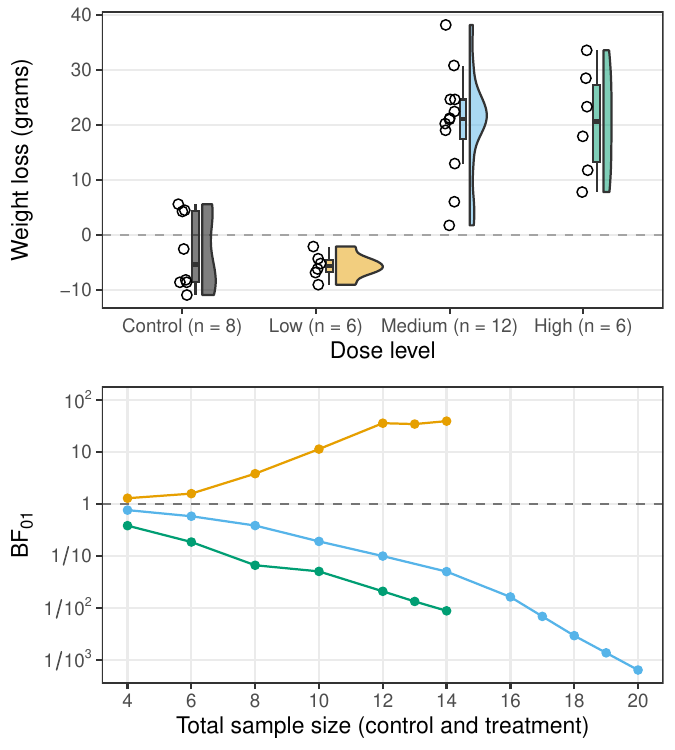} 

}

\end{knitrout}
\caption{Weight loss in rats assigned to different levels of a candidate drug
  \citep{Kang2025} (top plot). The bottom plot shows a sequential Bayes factor
  analysis contrasting the null hypothesis of no mean difference ($H_0 \colon
  \theta_i = 0$) to a mean difference of 5 grams ($H_0 \colon \theta_i =
  5$) for each treatment group $i \in
  \{\text{low},~\text{medium},~\text{high}\}$ indicated by the color. At each
  step an observation from control and treatment group is added until the
  maximum group size is reached.}
\label{fig:ratdata}
\end{figure}

In animal research, just as in clinical trials, every additional observation
carries significant ethical weight. Reducing sample size is hence of paramount
interest, which is one pillar of the 3R (``Replace, Reduce, Refine'') principles
in animal research \citep{RussellBurch1959}. We will now reanalyze data from a
preclinical experiment with rats, which was retrospectively analyzed by
Kang et al. \citep{Kang2025}. In the experiment, rats were randomly assigned to different
dose levels of a candidate drug or to a control group to estimate the drug's
effect on weight loss. Information that could be used to identify the drug or
study has been removed by Kang et al. due to confidentiality issues. The
top plot of Figure~\ref{fig:ratdata} shows the measured weight loss values
across groups. As can be seen, the low dose group (yellow) differs little from
the control group (black), whereas the medium (blue) and high dose (green)
groups show much higher weight loss.

In the analysis of Kang et al. \citep{Kang2025}, the effects of interest were the mean
differences in weight loss between the treatment groups and the control group,
denoted by $\theta_i$ with $i \in \{\text{low},~\text{medium},~\text{high}\}$.
The investigators defined a mean difference of at least 5 grams as effective,
which we will now use to specify the alternative hypothesis. The Bayes factor
contrasting the point hypotheses $H_0 \colon \theta_i = 0$ to $H_1 \colon
\theta_i = 5$ is shown in the bottom plot of Figure~\ref{fig:ratdata} for each
treatment group (color). An observation from each group is added at each step
until the maximum group size is reached. In each step, the mean difference
$\hat{\theta}_i = \widehat{\mathrm{E}}(Y_{i}) -
\widehat{\mathrm{E}}(Y_{\text{control}})$ and its standard error $\sigma_i =
\sqrt{\widehat{\mathrm{Var}}(Y_{i})/n_i +
  \widehat{\mathrm{Var}}(Y_{\text{control}})/n_{\text{control}}}$ are estimated
from the available data, which are then used to compute the Bayes factor from
Table~\ref{tab:BFcritical}. Note that the order in which the data were collected
is unknown, which is why a random permutation is shown here.
Appendix~\ref{app:rats} demonstrates that also for other permutations of the
data, similar results are obtained. As the sample size increases, the Bayes
factor in the low dose group (yellow) increases and reaches $\mathrm{BF}_{01} >
10$ after 10 rats, suggesting strong evidence for the absence of an effect over
its presence. In contrast, the Bayes factors in the medium (blue) and high
(green) dose groups decrease with increasing sample size, surpassing the
threshold for strong evidence of $\mathrm{BF}_{01} < 1/10$ after 12 and 8 rats,
respectively. Thus, if the experiment had been conducted using this sequential
Bayes factor design, $8 - \max\{10, 12, 8\}/2 = 2$
rats from the control group and $6 + 12 + 6 - (10 + 12 + 8)/2
=9$ rats from the treatment groups could have been
saved, totalling to 11 saved rats.

\begin{figure*}[!htb]
\begin{knitrout}
\definecolor{shadecolor}{rgb}{0.969, 0.969, 0.969}\color{fgcolor}

{\centering \includegraphics[width=\maxwidth]{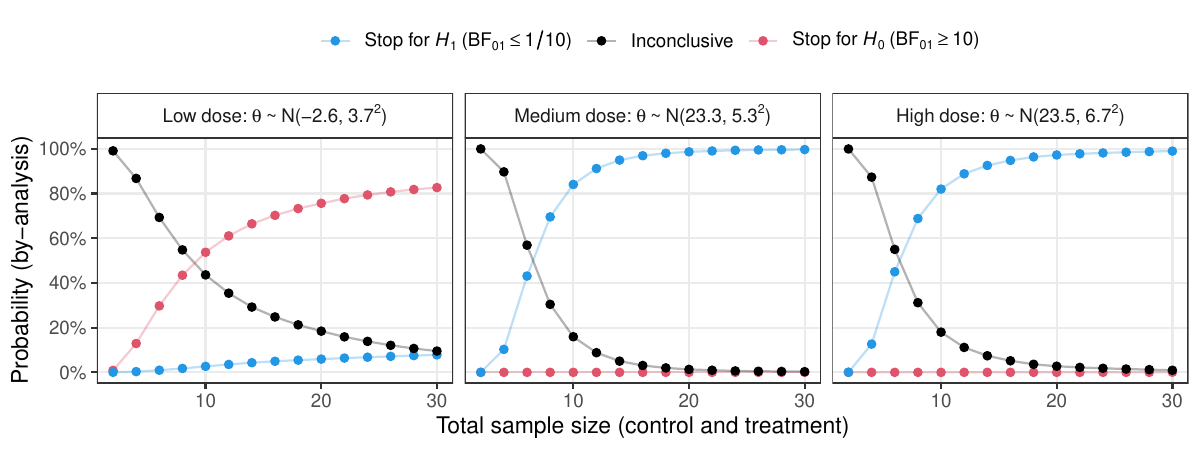} 

}

\end{knitrout}
\caption{Sequential Bayes factor probabilities for replications of weight loss
  rat experiments \citep{Kang2025}. Probabilities are computed assuming a design
  prior corresponding to posterior distributions for the mean difference
  $\theta$ based on data from the original experiments (indicated in the plot
  panels). Analyses are assumed to be conducted after each pair of rats from
  control and treatment groups.}
\label{fig:designrat}
\end{figure*}

Suppose that before moving on to trials with human participants, we want to
replicate these findings in independent experiments -- a practice which is
generally recommended in preclinical studies to rule out false positives
\citep{Piper2019, Drude2021}. In the Bayesian design of a replication study, it
is natural to use a design prior based on data from the original study to plan
the replication \citep{MicheloudHeld2022, Anderson2022, Pawel_etal2023}.
Figure~\ref{fig:designrat} shows stopping probabilities for replication study
designs in which an analysis is performed after each pair of rats from the
control and treatment groups. These designs assume a normal design prior
distribution for the mean difference $\theta$, centered around the estimated
mean difference from the original experiment, with a standard deviation equal to
the standard error of the estimate (see plot panels). Such a design prior can be
motivated as the posterior distribution of the mean difference based on the
original data and a flat prior. Note that if we would also use this design prior
as the analysis prior (instead of the point hypothesis $H_1 \colon \theta =
5$), a ``replication Bayes factor'' \citep{VerhagenWagenmakers2014,
  PawelHeld2022} would be obtained. However, here we will consider testing the
same alternative hypotheses as in the original study.

For the medium and high dose groups, the probabilities of finding strong
evidence for the alternative over the null hypothesis ($\mathrm{BF}_{01} \leq
1/10$; blue curves) quickly increase with increasing sample size. They
reach 80\% after a sample size of 10 in both cases. In contrast, for the low
dose group, the probability of finding strong evidence for the null over the
alternative hypothesis ($\mathrm{BF}_{01} \geq 10$; red curve) increases
more slowly, requiring nearly 30 rats to reach 80\%. While the probability of
misleading evidence remains nearly 0\% for the medium and high dose groups (red
curves), it increases considerably for the low dose group (up to almost 10\%,
blue curve) since the design prior based on the original data does not fully
rule out positive effects in the neighborhood of the alternative hypothesis
($H_1 \colon \theta = 5$).

In sum, the developed sequential Bayes factor design method enables rapid
calculation of key design characteristics while accounting for parameter
uncertainty. This can potentially lead to more efficient designs. For instance,
rather than allocating an equal number of rats to all groups, one could allocate
fewer rats to the medium and high dose groups and more rats to the low dose
group, thereby ensuring a high probability of informative inferences across all
groups.

\section{The sequential Bayesian \textit{t}-test}
\label{sec:ttest}

The Bayes factor version of the \textit{t}-test is a popular approach for
sequential hypothesis testing in the social sciences. Assuming normally
distributed data with unknown variance, Gronau et al. \citep{Gronau2020} proposed an
``informed'' \textit{t}-test Bayes factor that can take prior information into
account through informative prior distributions. The Bayes factor is
\begin{align}
  \mathrm{BF}_{01} = \frac{\mathrm{T}_{\nu}(t \mid 0, 1)_{(-\infty,+\infty)}}{\int_{-\infty}^{+\infty}
  \mathrm{NCT}_{\nu}(t \mid \theta \sqrt{n}) \, \mathrm{T}_{\kappa}(\theta \mid \mu, \tau)_{[a,b]}
  \, \mathrm{d}\theta}
  \label{eq:tBF}
\end{align}
where $t$ is the observed $t$-statistic, $n$ is the effective sample size (the
actual number of observations/pairs for one-sample/paired $t$-tests, or half the
harmonic mean of the group sizes for the two-sample $t$-tests), and
$\mathrm{T}_{\nu}(\cdot \mid \mu, \tau)_{[a,b]}$ is the density of the
location-scale $t$ distribution with location $\mu$, scale $\tau$, and
corresponding degrees of freedom $\nu$, truncated to the interval $[a, b]$.
$\mathrm{NCT}_{\nu}(\cdot \mid \lambda)$ is the density of the non-central $t$
distribution with non-centrality parameter $\lambda$. When the hyperparameters
are set to $\kappa = 1$, $\mu = 0$, $a = -\infty$, $b=+\infty$, the prior
becomes a Cauchy distribution, and the Bayes factor reduces to the widely used
``Jeffreys-Zellner-Siow'' (JZS) Bayes factor \citep{Jeffreys:1961, Zellner1980}.
The JZS Bayes factor is typically used with a scale of $\tau = 1/\sqrt{2}$, as
e.g., implemented in the \texttt{BayesFactor} R package \citep{Rouder2009}.

\begin{figure*}[!b]
\begin{knitrout}
\definecolor{shadecolor}{rgb}{0.969, 0.969, 0.969}\color{fgcolor}

{\centering \includegraphics[width=\maxwidth]{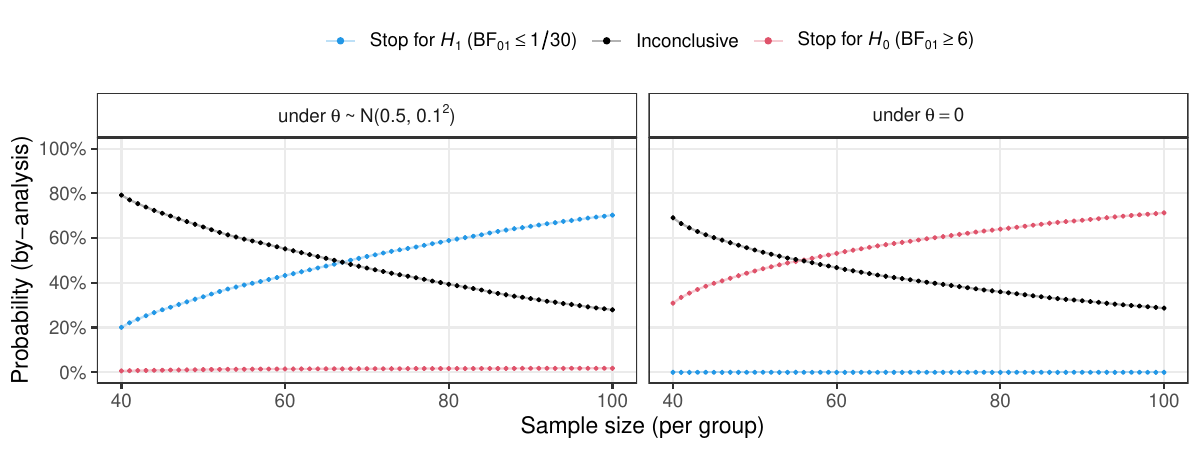} 

}

\end{knitrout}
\caption{Sequential Bayes factor design probabilities based on \textit{t}-test
  Bayes factor for design from Schönbrodt and Wagenmakers \citep{Schoenbrodt2018}.}
\label{fig:ttest}
\end{figure*}

Design calculations for the sequential Bayes factor \textit{t}-test can also be
embedded in the proposed \textit{z}-statistic framework. The Bayes
factor~\eqref{eq:tBF} is not available in closed-form but requires
one-dimensional numerical integration. Therefore, the critical \textit{t}-value
such that $\mathrm{BF}_{01} = k$ must be determined numerically, as demonstrated
by Pawel and Held \citep{Pawel2025} and Wong and Tendeiro \citep{Wong2025} for fixed designs. Assuming that the
effective sample size is large enough (e.g., $n \geq 30$), the \textit{t}
distribution can be approximated by a normal distribution. Specifically, for
large enough $n$, we have that $t \mid \theta \sim \mathrm{N}(\theta \sqrt{n},
1)$ where $\theta$ is the standardized mean (difference). Therefore, the vector
of accumulating \textit{t}-statistics approximately follows the canonical
distribution from Section~\ref{sec:canonical} with information level $I_i = n_i$
at analysis $i$, enabling the computation of stopping probabilities and other
design characteristics via numerical multivariate normal integration.

Schönbrodt and Wagenmakers \citep{Schoenbrodt2018} describe an application of
the \textit{t}-test Bayes factor to sequential experiments in psychology. They
consider an extreme design involving 61 analyses, in which an
analysis is performed after every additional pair of participants in treatment
and control groups until the maximum sample size $n_{61} =
100$ is reached or the experiment stopped before. They set asymmetric
Bayes factor thresholds $k_0 = 6$ and $k_1 = 1/30$ and
specify a normal design prior for the standardized mean difference $\theta \sim
\mathrm{N}(0.5, 0.1^2)$ to account for parameter uncertainty.
Finally, for the analysis they assume a default JZS Bayes factor with scale
$\tau = 1/\sqrt{2}$ truncated to positive standardized mean differences ($a =
0$, $b=+\infty$).

Figure~\ref{fig:ttest} shows the stopping probabilities for the design from
Schönbrodt and Wagenmakers \citep{Schoenbrodt2018}. Despite the many interim analyses, the calculations
took only a few seconds, whereas simulation-based calculations using the
\texttt{BFDA} package took around 20 minutes on the same computer. Under the
specified normal design prior (left plot), we obtain the stopping probabilities
for $H_1$ and $H_0$ as 70.3\% and
1.8\%, respectively, as well as an
average sample size of 69.4. These are very close to the
70.6\%, 1.6\%, and 69 reported by Schönbrodt and Wagenmakers, which were estimated
with simulation. Similarly, assuming no effect ($\theta = 0$, right plot), we
obtain 0\% to stop for $H_1$,
71.3\% to stop for $H_0$, and an
average sample size of 65.9, which are again very close to
the 0.6\%, 70.9\%, and 66 reported by Schönbrodt and Wagenmakers. In sum, the
\textit{z}-statistic perspective enables quick and accurate calculation of key
design characteristics of sequential designs based on the Bayes factor
\textit{t}-test, one of the most commonly used sequential Bayes factor tests.

\section{Discussion}
\label{sec:discussion}

Bayes factors are natural tools for sequential data analysis: Data can be
repeatedly analyzed without concerns about multiplicity, stopping decisions are
naturally linked to interpretable updates of prior to posterior odds of
competing hypotheses, and, unlike posterior tail probabilities, there is no need to
specify prior probabilities for competing hypotheses. Despite these advantages,
the broader adoption of sequential Bayes factor designs has been limited, one
potential reason being the difficulty of computing their design characteristics.
Existing approaches typically rely on extensive simulation, which can be
computationally costly, sensitive to Monte Carlo error, and inconvenient when
exploring many design options.

In this paper, we introduced a general approach that overcomes these limitations
by expressing Bayes factors as functions of \textit{z}-statistics and
extending results from classical group sequential design theory. This
perspective showed that Bayes factor stopping rules correspond to sets of
hyper-rectangle regions in the space of accumulating \textit{z}-statistics.
Under the canonical \textit{z}-statistic distribution, the probability of these
regions can be computed efficiently using multivariate normal integration,
eliminating the need for simulation. The resulting computations are fast,
accurate, and scalable to designs with many interim looks. The approach also
naturally incorporates design priors, enabling experiments to account for
parameter uncertainty at the design stage. Traditional fixed parameter
calculations are a special case, thus enabling flexible exploration of both
Bayesian and frequentist operating characteristics within a unified framework.

There are, however, some limitations. The method requires that the Bayes factor
can be expressed as a function of the \textit{z}-statistic and that the
accumulating \textit{z}-statistics follow, at least approximately, a canonical
multivariate normal distribution. Situations where these assumptions are not met
(e.g., binary data with small sample size and/or extreme probabilities) may
still require simulation or other numerical methods for computing design
characteristics. Developing modifications for these situations could be one
avenue for future research.

To conclude, many sequential Bayes factor designs can be planned as rapidly and
reliably as classical group sequential designs. Our accompanying R package
\texttt{bfpwr} implements these methods and offers experimenters a practical
tool for designing efficient and informative studies.

\subsection*{Acknowledgments}
We thank Wong Tsz Keung, Riko Kelter, and František Bartoš for valuable comments
on drafts of the manuscript. We thank Torsten Hothorn for help with calculating
multivariate normal probabilities with \texttt{mvtnorm::lpmvnorm}. We thank Tony
Pourmohamad for pointing us to the data from the rat experiment. The
acknowledgment of these individuals does not imply their endorsement of the
paper. We used Claude \citep{Claude2026} for language and grammar checking and
for reviewing and debugging R code during manuscript preparation. We reviewed,
revised, and approved all AI-assisted content and take full responsibility for
the final manuscript.

\subsection*{Conflict of interest}
We declare no conflict of interest.

\subsection*{Software and data}
Data from the Low-PV trial were extracted from Table S5 in the supplement of
Barbui et al. \citep{Barbui2021}. Data from the weight loss rat experiment were extracted from
Figure~ 4 in Kang et al. \citep{Kang2025}. Code and data to reproduce our analyses are
openly available at \url{https://github.com/SamCH93/bfgsd}. A snapshot of the
repository at the time of writing is available at
\url{https://doi.org/10.5281/zenodo.18160652}. We used the statistical
programming language R version 4.6.0 (2026-04-24) for analyses
\citep{R} along with the \texttt{ggplot2} \citep{Wickham2016}, \texttt{dplyr}
\citep{Wickham2023}, \texttt{mvtnorm} \citep{Genz2009}, \texttt{ggpubr}
\citep{Kassambara2023}, \texttt{ggrain} \citep{Allen2021}, \texttt{xtable}
\citep{Dahl2019}, \texttt{rpact} \citep{WassmerBrannath:2016}, and
\texttt{knitr} \citep{Xie2015} packages.


\begin{appendix}

\section{The R package bfpwr}
\label{app:package}

The following code excerpt shows how the \texttt{bfpwr} R package can be used to
compute design characteristics of a sequential JZS (\textit{t}-test) Bayes
factor design.


\begin{knitrout}\scriptsize
\definecolor{shadecolor}{rgb}{0.969, 0.969, 0.969}\color{fgcolor}\begin{kframe}
\begin{alltt}
\hlcom{## group sequential design features not in CRAN version yet}
\hlcom{## remotes::install_github(repo = "SamCH93/bfpwr", subdir = "package",}
\hlcom{##                         ref = "gsd")}
\hlkwd{library}\hldef{(bfpwr)} \hlcom{# load package}
\hlcom{## set up sequential t-test Bayes factor design}
\hldef{design} \hlkwb{<-} \hlkwd{ptbf01seq}\hldef{(}
    \hlkwc{k1} \hldef{=} \hlnum{1}\hlopt{/}\hlnum{10}\hldef{,} \hlcom{# Bayes factor threshold for H1}
    \hlkwc{k0} \hldef{=} \hlnum{6}\hldef{,} \hlcom{# Bayes factor threshold for H0}
    \hlkwc{type} \hldef{=} \hlsng{"two.sample"}\hldef{,} \hlcom{# two-sample t-test}
    \hlkwc{n} \hldef{=} \hlkwd{seq}\hldef{(}\hlnum{20}\hldef{,} \hlnum{100}\hldef{,} \hlnum{20}\hldef{),} \hlcom{# per-group sample sizes at analyses}
    \hlcom{## specify one-sided Jeffreys-Zellner-Siow analysis prior}
    \hlkwc{plocation} \hldef{=} \hlnum{0}\hldef{,} \hlkwc{pscale} \hldef{=} \hlnum{1}\hlopt{/}\hlkwd{sqrt}\hldef{(}\hlnum{2}\hldef{),}
    \hlkwc{pdf} \hldef{=} \hlnum{1}\hldef{,} \hlkwc{alternative} \hldef{=} \hlsng{"greater"}\hldef{,}
    \hlcom{## specify normal design prior around SMD = 0.5}
    \hlcom{## with small stand. deviation}
    \hlkwc{dpm} \hldef{=} \hlnum{0.5}\hldef{,} \hlkwc{dpsd} \hldef{=} \hlnum{0.05}
\hldef{)}
\hldef{design} \hlcom{# print design summary}
\end{alltt}
\begin{verbatim}
## 
## Sequential Bayes Factor Design
## --------------------------------
## H0:               SMD (stand. mean diff.)  = 0
## H1:               SMD (stand. mean diff.)  > 0
## Analysis prior:   SMD|H1 ~ t(location = 0, scale = 0.7071, df = 1)_+
## Design prior:     SMD ~ N(mean = 0.5, sd = 0.05)
## BF thresholds:    H1 if BF01 <= 1/10, H0 if BF01 >= 6
## Number of looks:  5
## Sample sizes 1:   20, 40, 60, 80, 100
## Sample sizes 2:   20, 40, 60, 80, 100
## 
## 
## Stagewise cumulative probabilities:
##  Stage Pr(H1 stop) Pr(H0 stop) Pr(inconclusive)
##      1      0.1302      0.0041           0.8656
##      2      0.3499      0.0070           0.6430
##      3      0.5497      0.0082           0.4421
##      4      0.7017      0.0087           0.2897
##      5      0.8069      0.0088           0.1843
## 
## Expected sample size 1: 64.8085
## Expected sample size 2: 64.8085
## Standard deviation of sample size 1: 28.3782
## Standard deviation of sample size 2: 28.3782
## 
## NOTE:  BF01 < 1 indicates evidence for H1 over H0
\end{verbatim}
\begin{alltt}
\hlkwd{plot}\hldef{(design)} \hlcom{# plot design prior (top) and under H0 (bottom)}
\end{alltt}
\end{kframe}

{\centering \includegraphics[width=\maxwidth]{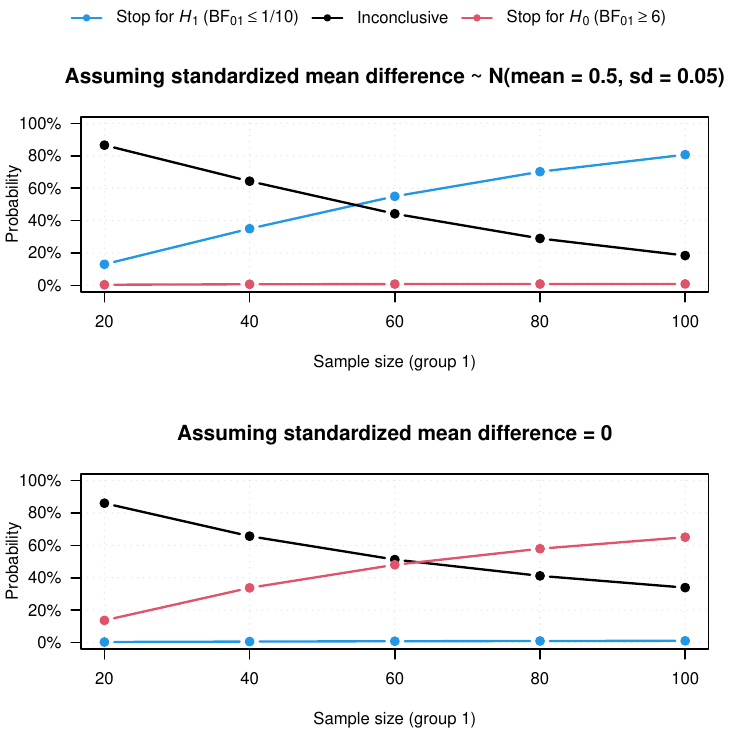} 

}

\end{knitrout}


\section{Marginal distribution of the \textit{z}-statistics}
\label{app:zmarginal}
The \textit{z}-statistic vector can be represented as $\boldsymbol{Z} \mid \theta
= \theta \boldsymbol{I} + \boldsymbol{\epsilon}$ with $\boldsymbol{\epsilon}
\sim \mathrm{N}_m(\boldsymbol{0}, \boldsymbol{\Sigma})$ and independent of
$\theta$. Since $\boldsymbol{I}$ is fixed and also $\theta \sim
\mathrm{N}(\mu_d, \tau^2_d)$, the marginal distribution of $\boldsymbol{Z}$ is
also normal. By the law of total expectation, its expectation is
\begin{align*}
  \mathrm{E}(\boldsymbol{Z})
  &= \mathrm{E}\{\mathrm{E}(\boldsymbol{Z} \mid \theta)\} \\
  &= \mathrm{E}(\theta \boldsymbol{I}) \\
  &= \mathrm{E}(\theta)  \boldsymbol{I} \\
  &= \mu_d  \boldsymbol{I}.
\end{align*}
Similarly, applying the law of total covariance, its covariance is
\begin{align*}
  \mathrm{Cov}(\boldsymbol{Z})
  &= \mathrm{E}\{\mathrm{Cov}(\boldsymbol{Z} \mid \theta)\} + \mathrm{Cov}\{\mathrm{E}(\boldsymbol{Z} \mid \theta)\} \\
  &= \mathrm{E}(\boldsymbol{\Sigma}) + \mathrm{Cov}(\theta \boldsymbol{I}) \\
  &= \boldsymbol{\Sigma} + \boldsymbol{I} \, \mathrm{Cov}(\theta) \boldsymbol{I}^\top \\
  &= \boldsymbol{\Sigma} + \tau^2_d \boldsymbol{I} \boldsymbol{I}^\top. \\
\end{align*}


\section{Sensitivity analysis for rat experiment}
\label{app:rats}
\begin{figure}[!htb]
\begin{knitrout}
\definecolor{shadecolor}{rgb}{0.969, 0.969, 0.969}\color{fgcolor}

{\centering \includegraphics[width=\maxwidth]{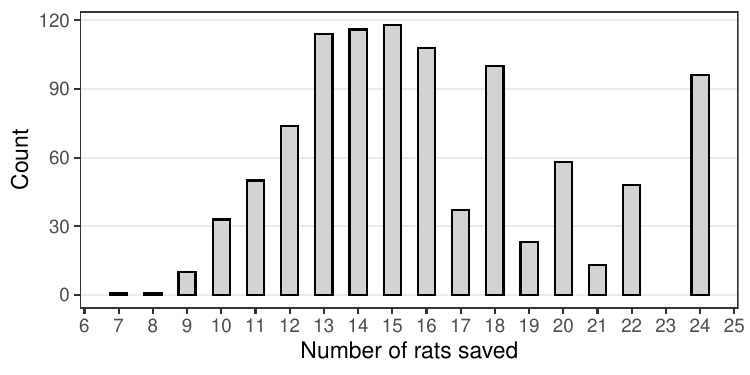} 

}

\end{knitrout}
\caption{Number of rats saved across 1'000
  random permutations of the data order when using a sequential Bayes factor
  design.}
\label{fig:ratspermute}
\end{figure}

Figure~\ref{fig:ratspermute} and Table~\ref{tab:kangpermute} show results from
sensitivity analyses regarding the random permutation of the data set from
Kang et al. \citep{Kang2025}, for which the original collection order is unknown. The
11 rats saved reported in Section~\ref{sec:rat} are
representative, or even a conservative estimate compared to the the permutation
distribution in Figure~\ref{fig:ratspermute}.
Similary, the decisions to stop for $H_0$ in the low dose and to stop for $H_1$
in the other groups reported in Section~\ref{sec:rat} corresponds to the
majority of decisions across the permutations (Table~\ref{tab:kangpermute}).
\begin{table}[!htb]
  \centering
  \caption{Proportion of sequential Bayes factor stopping decisions for $H_{0}$
    or $H_{1}$
    across 1'000 random permutations of the
    data order.}
  \label{tab:kangpermute}
\begin{tabular}{llll}
  \toprule
Treatment group & Stop for $H_0$ & Stop for $H_1$ & Indecisive \\ 
  \midrule
Low & 89.5\% & 10.5\% & 0\% \\ 
  Medium & 0\% & 100\% & 0\% \\ 
  High & 0\% & 100\% & 0\% \\ 
   \bottomrule
\end{tabular}

\end{table}

\end{appendix}

\bibliographystyle{nessart-number}
\bibliography{bibliography}



\end{document}